
\magnification\magstep1
\font\cs=cmr10 scaled \magstep3
\vskip 1.5 true cm
\centerline{\cs Conservation laws}
\vskip 0.7 true cm
\centerline{\cs for}
\vskip 0.7true cm
\centerline{\cs Cosmological Perturbations}
\vskip 1 true cm
\centerline{Nathalie Deruelle$^{1,2}$ and Jean-Philippe Uzan$^1$}
\smallskip
\vskip 1 true cm
\centerline{$^1$ D\'epartement d'Astrophysique Relativiste et de
Cosmologie, }
\centerline{UPR 176 du Centre National de la Recherche Scientifique,}
\centerline{Observatoire de Paris, 92195 Meudon, France}
\bigskip
\centerline{$^2$ DAMTP, University of Cambridge,}
\centerline{Silver Street, Cambridge, CB3 9EW, England}

\vskip 1cm

\centerline{Peyresq Proceedings, 1997}

\vskip 2cm

\noindent
{\bf Abstract}
\bigskip
We briefly recall the problem of defining conserved quantities such as
energy in general relativity, and the solution given by introducing a
symmetric background. We apply the general formalism to perturbed
Robertson-Walker space-times with de Sitter geometry as background. We
relate the obtained conserved quantities to Traschen's integral
constraints and mention a few applications in cosmology.
\vskip 1cm

\noindent
{\bf 1. Introduction}
\bigskip
Consider Maxwell's equations~: $$D_\nu F^{\mu\nu}=\mu_0j^\mu\eqno(1)$$
where the electromagnetic tensor is $F_{\mu\nu}=\partial_\mu A_\nu
-\partial_\nu A_\mu$, $A_\mu$ being the vector potential, where
$j^\mu$ is the electromagnetic current, where $\mu_0$ is a coupling
constant between the current and the field it creates, and where
$D_\mu$ is the covariant derivative associated with the metric
$g_{\mu\nu}$ (with determinant $g$). Since $F_{\mu\nu}$ is
antisymmetric they can be rewritten as~:
$$\partial_\nu(\sqrt{-g}F^{\mu\nu})=\mu_0\sqrt{-g}j^\mu.\eqno(2)$$
They therefore yield a conservation law and hence an integral
equation. Indeed we have, applying Gauss' theorem~:
$$\partial_\mu(\sqrt{-g}j^\mu)=0\qquad \Rightarrow\qquad\partial_0
\int_V\sqrt{-g}j^0dV=-\int_{\partial V}
\sqrt{-g}j^idS_i.\eqno(3)$$  If the volume $V$ is taken
to be the whole space and if there are no currents on the boundary
$\partial V$ then we have that the total charge $e$ defined as
$$e\equiv{1\over\mu_0}\int_V\sqrt{-g} j^0dV\eqno(4)$$ is constant~:
$\partial_0 e=0$. Using Maxwell's equations (2) we can moreover
express it as a surface integral~: $$e={1\over\mu_0}\int_{\partial
V}\sqrt{-g}F^{0i}dS_i.\eqno(5)$$

On another hand one can construct a tensor, the stress-energy tensor~:
$$-\mu_0T_{\mu\nu}\equiv
F_{\mu\rho}F^\rho{}_\nu+{1\over4}g_{\mu\nu}F^{\rho\sigma}F_{\rho\sigma}\eqno(6)$$
which, thanks to Maxwell's equations again, is such that~: $$D_\mu
T^{\mu\nu}+F^\nu{}_\mu j^\mu=0.\eqno(7)$$ Outside the charges and in
Minkowski space-time in cartesian coordinates, this equation is also a
conservation law and yields another integral equation~: $$\partial_\mu
T^{\mu\nu}=0\quad
\Rightarrow\partial_0\int_VT^{0\nu}dV=-\int_{\partial V}T^{i\nu}dS_i.\eqno(8)$$
 If the field decreases fast
enough at infinity the cartesian vector
$P^\nu$ defined by~:
$$P^\nu\equiv \int_VT^{0\nu}dV\eqno(9)$$
 is therefore constant~:
$\partial_0P^\nu=0$.

These conservation laws and integral equations (3) and (8) are mere
consequences of Maxwell's equations. In other words, given charges in
arbitrary motion, the field $F_{\mu\nu}$ they create is such that the
total charge defined by (4) and, cartesian coordinates being used, the
energy-momentum vector defined by (9), are constant if the boundary
terms are zero. In addition the charge is also given by the surface
integral (5).

\vskip 0.5cm

Now it is well-known that those laws and equations reflect in fact the
symmetries of the theory.  Indeed the conservation of the charge,
equation (3a), follows from the requirement that, like Maxwell's
equations, the action they derive from~:
$$S\equiv\int\sqrt{-g}Ld^4x\qquad\hbox{with}\qquad
L=\sqrt{-g}\left(j^\mu
A_\mu-{1\over4\mu_0}F_{\mu\nu}F^{\mu\nu}\right)\eqno(10)$$ be the same
(up to a total derivative) in a gauge transformation~: $A_\mu\to
A_\mu-\partial_\mu f$.

As for the conservation of the stress-energy tensor (8a), it reflects
the homogeneity and isotropy of Minkowski spacetime. Indeed, in
cartesian coordinates the action (10) does not depend explicitely on
the coordinates $x^\mu$, thus Computing $\partial_\mu L$, using
Maxwell's equations, then yields a tensor which, after symmetrisation,
is nothing but (6) (in cartesian coordinates), and is conserved
(Noether's theorem).

This (symmetric) stress-energy tensor $T_{\mu\nu}$, can also be
defined as the functional derivative of the action with respect to the
metric, and equation (7) follows from the fact that the action is a
scalar.  Minkowski space-time being maximally symmetric, it possesses
10 Killing vector $\xi_\mu$, such that
$D_\mu\xi_\nu+D_\nu\xi_\mu=0$. Hence outside the charges, equation (7)
yields~: $$\partial_\mu(\sqrt{-g}T^\mu{}_\nu\xi^\nu)=0,\eqno(11a)$$
which is the generalisation to any coordinate system of the
conservation law (8a). The integral equation which ensues is~:
$$\partial_0P(\xi)=-\int_{\partial V}\sqrt{-g}T^{i\nu}\xi_\nu
dS_i\quad\hbox{where}\quad P(\xi)\equiv\int_V\sqrt{-g}T^0{}_\nu\xi^\nu
dV.\eqno(11b)$$ In cartesian coordinates and for the four $\xi_\mu$
corresponding to time or space translations, $P(\xi)$ is the vector
defined by (9) whose constancy (for an isolated system) therefore
reflects the symmetries of Minkowski spacetime.

\vskip 0.5cm

In General Relativity, first, gauge invariance and invariance under
coordinate transformations are one and the same thing so that the
notions of ``charge" and ``energy-momentum" of the gravitational
system coalesce.

Second, Einstein's equations~: $$G_{\mu\nu}=\kappa
T_{\mu\nu},\eqno(12)$$ where $G_{\mu\nu}$ is Einstein's tensor,
$\kappa=8\pi G/c^4$ Einstein's coupling constant and $T_{\mu\nu}$ the
stress-energy tensor of matter (defined as the functional derivative
of the matter action with respect to the metric), imply, via Bianchi
identity~: $$D_\mu T^{\mu\nu}=0.\eqno(13)$$

Equations (13) (which are the gravitational analogue of equation (7))
can be manipulated in various ways to yield conservation laws (similar
to equations (3a) or (8a)).  Landau and Lifchitz (1962) rewrote them
as~: $$\partial_\mu[(-g)(T^{\mu\nu}+t^{\mu\nu}_{LL})]=0\eqno(14)$$
where $t^{\mu\nu}_{LL}$ is some expression quadratic in the
Christoffel symbols. Hence the quantity~:
$$P^\nu_{LL}=\int_V(-g)(T^{0\nu}+t^{0\nu}_{LL})dV\eqno(15)$$ is
constant if surface terms vanish. It is the gravitational analogue of
the charge (4) or the electromagnetic energy-momentum vector (9). And,
like the charge (see equation (5)), it can be expressed as a surface
integral, in terms of a ``super-potential" from which the
pseudo-tensor $t^{0\nu}_{LL}$ derive.  However $P^\nu_{LL}$ is not a
vector under general coordinate transformations and moreover has the
wrong tensorial weight. Hence it does not really qualify as a proper
definition of energy-momentum.

Einstein on the other hand applied in 1915 Noether's theorem to the
Hilbert action from which his equations derive. Indeed, as a
functional of the metric it does not depend explicitely on the
coordinates. He therefore also obtained conservation laws~:
$$\partial_\mu[\sqrt{-g}(T^{\mu}{}_{\nu}+t^{\mu}_{E\nu})]=0\eqno(16)$$
where $t^{\mu}_{E\nu}$ is yet another expression quadratic in the
Christoffel symbols.  Hence the quantity~:
$$P^\nu_{E}=\int_V\sqrt{-g}(T^{0\nu}+t^{0\nu}_E)dV\eqno(17)$$ is also
constant if surface terms vanish, and can also be written as a surface
integral. Despite the fact that it is obtained from Noether's theorem
and therefore reflects some properties of spacetime and has the right
tensorial weight, (17) does not qualify either as a proper definition
of energy-momentum as it is not a vector under general coordinate
transformations. Moreover, $t^{\mu\nu}_E\equiv
g^{\mu\rho}t^\nu_{E\rho}$ is not symmetric and cannot define an
angular momentum (this is in fact this problem which led Landau and
Lifchitz to (14-15)).
\vskip0.5cm

As advocated by many authors (see [1-2] for reviews), a possible way
out of this problem of defining energy, momentum (and angular
momentum) in General Relativity is to introduce a background
spacetime. We now briefly summarize this approach, following reference
[3].

\bigskip
\noindent
{\bf 2. Defining energy etc with respect to a background}
\bigskip

Consider a spacetime $({\cal M}, g_{\mu\nu}(x^\lambda))$, a
background $(\bar{\cal M}, \bar g_{\mu\nu}(x^\lambda))$ and a mapping
between these two spacetimes.

Take as lagrangian density for gravity
$$ {\hat {\cal L}}_{G}={1\over2\kappa}[{\hat g}^{\mu\nu}
(\Delta^\rho_{\mu\nu}\Delta^\sigma_{\rho\sigma} -
\Delta^\rho_{\mu\sigma}\Delta^\sigma_{\rho\nu}) -({\hat g}^{\mu\nu}
-\bar{\hat g}^{\mu\nu}){\bar R}_{\mu\nu}]
\eqno(18)$$
where we have
introduced the difference $\Delta^\lambda_{\mu\nu}$ between Christoffel
symbols in ${\cal M}$ and ${\bar{\cal M}}$ and where $\bar R_{\mu\nu}$
is the Ricci tensor of the background. A hat
denotes multiplication by $\sqrt{-g}$. Since the ``$\Delta$" are tensors,
$\hat {\cal L}_{G}$ is a true scalar density.  

If we now perform a small
displacement $\Delta x^\mu = \zeta^\mu\Delta\lambda$, where
$\zeta^\mu$ is an arbitrary vector field and $\Delta\lambda$ an
infinitesimal parameter, and use the fact that $\hat {\cal L}_{G}$ is a
scalar density, we have that, with ${\rm L}_\zeta$ denoting the Lie
derivative,
$$ {\rm L}_\zeta \hat {\cal L}_G
-\partial_\mu(\hat {\cal L}_G \zeta^\mu)=0.
\eqno(19)$$
Computing explicitely ${\rm L}_\zeta \hat {\cal L}_G$ from (18), it can
be shown (cf [2]) that there exists an identically conserved vector
$\hat I^\mu$, analogous to the electromagnetic current, such that 
$$\partial_\mu \hat I^\mu= 0\eqno(20a)$$
yielding the integral equation~:
$$\partial_0P(\zeta)=-\int_{\partial V}\hat I^i
dS_i\quad\hbox{where}\quad P(\zeta)\equiv\int_V\hat I^0 dV.\eqno(20b)$$
Equations (20a-b) are the gravitational analogue of equations (3) and (11a-b). 

Now it also follows from (20a) that there exists
 an antisymmetric tensor $\hat J^{[\mu\nu]}$ such that
$$ \hat I^\mu = \partial_\nu \hat J^{[\mu\nu]}.
\eqno(21)$$
This is the gravitational analogue of Maxwell's equation (2). Hence,
just like the electric charge (see equation (5)), $P(\zeta)$ can be
expressed as a surface integral~: $$P(\zeta)=\int_{\partial V}\hat
J^{0i}dS_i.\eqno(22)$$

The explicit expressions for $\hat I^\mu$  and
for
$J^{[\mu\nu]}$ can be found in [2] (see also [3]).  

The equalities (20-22) are valid for all
$\{g^{\mu\nu},\bar g^{\mu\nu}, \zeta^\nu\}$.  
They become the Noether conservation laws when the vectors $
\zeta^\mu$ are Killing vectors of the background. 
Therefore, in order to obtain the maximum number of Noether
conservation laws, one is led to consider a background with maximal
symmetry, in which case ten integral equations (one for each Killing
vector) can be written.  If the Killing vector refers to the time
translations in Minkowski spacetime or the quasi-time translations of
de Sitter spacetime, then the corresponding quantity $P(\zeta)$ will
be called energy.  When one uses the three Killing vectors associated
with the Lorentz rotations of Minkowski or the quasi-Lorentz rotations
of de Sitter spacetimes, $P(\zeta)$ will be the ``position of the
centre of mass", etc.  The introduction of a mqaximally symmetric
background thus allows to define an energy etc, even if the physical
spacetime does not possess symmetries, globally or asymptotically.
The justification for such a terminology can be found in e.g. [2].

\bigskip
\noindent
{\bf 3. The energy of a cosmological perturbations with respect to 
de Sitter space}
\bigskip
We now apply the formalism
summarized above to a perturbed Robertson-Walker spacetime with metric~:
$$ds^2=dt^2-a^2(t)(f_{ij} + h_{ij})dx^idx^j \eqno(23)$$
  $f_{ij}$ is the metric of a 3-sphere, plan
or hyperboloid depending on whether the index $k=(+1,0,-1)$~:
$$ f_{ij}=\delta_{ij}+k{\delta_{im}\delta_{jn}x^mx^n\over{1-kr^2}}
\qquad\hbox{with}\qquad r^2\equiv\delta_{ij}x^ix^j.\eqno(24)
$$ The scale factor $a(t)$ is determined by Friedmann's equation and
$h_{i j }(x^\mu)$ is a small perturbation of $f_{ij}$ in a synchronous
gauge.  The maximally symmetric background is chosen to be de Sitter
space with the same spatial topology as the physical perturbed
Robertson-Walker spacetime and its metric will be written as~: $$
d\bar s^2=\Psi(t)^2 dt^2 - \bar a(t)^2f_{ij}dx^idx^j
\eqno(25)$$
Equation (25) contains a definition of
the mapping for each point of the $t=Const.$ hypersurface, up to an
isometry.  The function $\Psi(t)$ defines the mapping of the cosmic
times (and the explicit expression for the scale factor $\bar a(t)$).

The explicit expressions of the ten de Sitter Killing vectors when the
metric is written under the form (25) can be found in e.g. [3].

The zeroth order conserved quantities $P_{RW}(\zeta)$ have been
defined and studied by Katz Bi$\check {\rm c}$ak and Lynden-Bell
[2]. Their perturbations at first order were given in [3]. The final
result is~: $P(\zeta)=P_{RW}(\zeta) +\delta P(\zeta)$, with~: $$
\delta P(\zeta) \equiv\int_V
\sqrt{-g}\left(\delta T^0_\mu\zeta^\mu +{1\over2} \beta{\dot{\tilde
h}}\zeta^0\right)dV+ \int_{\partial V}\hat M^ldS_l= \int_{\partial V}
(\hat B^l+\hat M^l)dS_l \eqno(26) $$ where we have introduced the
notations $\kappa\beta \equiv {\dot a}/{a} - {\dot {\bar a}}/{\bar a}
$ and $\tilde h\equiv -2f^{ij}h_{ij}$, and where the explicit
expressions of the $zeta$-dependent surface terms $M^l$ and $B^l$ can
be found in [3].

Using the explicit expressions of the De Sitter/Robertson-Walker
Killing vectors corresponding to spatial translations,
$\zeta^\mu=P^\mu$, the total momentum of the perturbations is thus
defined as $$ \delta P_i(P)\equiv a^3
\int_V dV\delta T^0_i+\int_{\partial V}\hat
M^l_idS_l=\int_{\partial V}(\hat B^l_i+\hat M^l_i)dS_l. \eqno(27) $$
Hence the total momentum is the sum of a background and mapping
independent volume integral plus a surface term which does depend on
the background and the mapping.  The same holds for the total angular
momentum.

When it comes now to the de Sitter Killing vectors corresponding to
quasi-time translations ($\zeta^\mu=T^\mu$) and quasi-Lorentz
rotations ($\zeta^\mu= K^\mu$), equations (26) can be written under
the form~: $$\delta P(T) = {1\over\Psi}\delta P_{Tr}(T)
+\int_{\partial V}f ({ \hat M}^l +\hat C^l)dS_l \eqno(28) $$ $$
\left\lbrace \matrix{
\delta P^i( K) ={1\over\Psi}\delta P^i_{Tr}(K) + \int_{\partial V}({\hat
M}^{li} +{\hat D}^{li})dS_l
\hfill & {\rm for} \;  k\neq0 \cr 
\delta P^i(K) = {1\over\Psi}\delta
P^i_{Tr}(K) - {1\over2\bar H\bar a^2}\delta P^i(P) +\int_{\partial V}
({\hat M}^{il} +{\hat E}^{il})dS_l & {\rm for} \;  k=0 \cr }\right.
\eqno(29)$$
where~:
$$ \delta
P_{Tr}(T)\equiv a^3 \int_V \left(\delta\rho-H\delta T^0_lx^l\right)
dV=\int_{\partial V}\hat B^l(T) dS_l  \eqno(30)
$$
$$ \left\lbrace
\matrix{ \delta P^i_{Tr}(K)\equiv a^3 \int_V\left[x^i\delta\rho
+H\delta
T^0_l\left(k\delta^{li}-x^lx^i\right)\right]{dV\over\sqrt{1-kr^2}}=
\int_{\partial V}\hat B^{li}(K) dS_l \; & {\rm for} \; k\neq0 \cr \delta
P^i_{Tr}(K)\equiv a^3\int_V\left[x^i\delta\rho +H\delta T^0_l\left(
{1\over2}\delta^{li}r^2-x^lx^i\right)\right]dV=\int_{\partial V}\hat
B^{li}(K)dS_l \; & {\rm for} \; k=0 \cr }\right.\eqno(31)$$ and where
the explicit expressions for the various surface terms can be found in
[3].

Hence, the energy and motion of the centre of mass of the perturbations
are also the sum of volume integrals which are, up to the overall function of
time $\Psi$, background and mapping independent, plus surface terms which
do depend on the background and the mapping. 

Turning to localised perturbations for which all surface integrals
vanish, we see on the form (27-31) for the conserved quantities that
the resulting constraints are background and mapping independent. As
shown in [3] they are equivalent to Traschen's constraints [4], which
have been widely used for treating localised perturbations (also
called ``causal" or ``active") (see e.g. ref [5-10]).

\bigskip
\noindent
{\bf Acknowledgements}
\bigskip
The original work on which this short review is based upon was done
with Joseph Katz. We thank him for the most enjoyable
collaboration. We also thank Mrs Smet for her warm hospitality in
Peyresq.

\bigskip
\noindent
{\bf References}
\bigskip
[1] Katz J 1996 in Gravitational Dynamics edited by Lahav,
Terlevitch and Terlevitch (Cambridge University Press, p.  193)

[2]
{Katz J, Bi${\check {\rm c}}$ak J, Lynden-Bell D, 1996}, to be published in 
{Phys. Rev. D.}

[3] N. Deruelle, J. Katz, J.P. Uzan, 1997, Class. and Quantum Grav. 14, 1

{[4]} {Traschen J 1984}, {Phys. Rev. D}, {29}, {1563}

{[5]} {Abbott L F,
Traschen J, Xu Rui-Ming 1988}, {Nucl. Phys.}, {B296}, {710}

{[6]} {Traschen J 1985}, {Phys. Rev. D}, {31}, {283}

{[7]} {Traschen J and Eardley D 1986}, {Phys. Rev. D}, {34}, {1665}

{[8]} {Veeraraghavan S and Stebbins A 1990}, {Astrophysical Journal}, {365}, 
{37-65}

{[9]} {Traschen J, Turok N and Brandenberger R 1986}, {Phys. Rev. D}, {34}, 
{919}

[10] J.P. Uzan, N. Turok, ``Using Conservation Laws to Study Cosmological 
Perturbations in Curved Universes", to appear in Phys. Rev. {\bf D57}.

\end